\documentclass[twocolumn,printnumbers,amsmath,amssymb]{revtex4}
\usepackage{graphicx}
\usepackage{color}
\usepackage{pdfpages}

\begin{document}

\title{ Low-frequency excess vibrational modes in two-dimensional glasses }

\date{\today}

\author{Lijin Wang$^{1,*}$}
\author{Grzegorz Szamel$^{2}$}
\author{Elijah Flenner$^{2,\dag}$}

\affiliation{$^1$School of Physics and Materials Science, Anhui University, Hefei 230601, P. R. China}

\affiliation{$^2$Department of Chemistry, Colorado State University, Fort Collins, Colorado 80523, USA}

\begin{abstract}

Glasses possess more low-frequency vibrational modes than predicted by Debye theory. These excess modes
are crucial for the understanding of the low temperature thermal and mechanical properties of glasses,
which differ from those of crystalline solids. Recent simulational studies suggest that the density of the
excess modes scales with their frequency $\omega$ as $\omega^4$ in two and higher dimensions.
Here, we present extensive  numerical studies of two-dimensional model glass formers
over a large range of glass stabilities. We find that the density of the excess modes follows
$D_\text{exc}(\omega)\sim \omega^2 $  up to around the boson peak, regardless of the glass stability.
The stability dependence of the overall scale of $D_\text{exc}(\omega)$ correlates with
the stability dependence of low-frequency sound attenuation.
However, we also find that in small systems, where the
first sound mode is pushed to higher frequencies,  at frequencies below the first sound mode
there are excess modes with a system size independent density of states that scales as $\omega^3$.
\end{abstract}

\maketitle

Low-temperature glasses exhibit thermal and mechanical
properties~\cite{Zeller1971,Anderson1972,Zaitlin1975,Phillips,Pohl2002,Ozawa_PNAS2018, Ketkaew_2018NC}
that distinguish them from low temperature crystalline solids. The low frequency vibrational modes
in crystalline solids are plane waves. Their density of states is well described by Debye theory and scales with
frequency $\omega$ as $\omega^{d-1}$ where $d$ is the spatial dimension.
For glasses, there are additional low-frequency modes that result in a  peak in the reduced total density of states
$D(\omega)/ \omega^{d-1}$ in $d$ spatial dimensions, which is referred to as
the boson peak~\cite{parisi_nature,Inoue1991,Buchenau2007,tanaka_nm2008}. Understanding the nature of
the additional modes provides insight into the the physics behind the anomalous properties
of glasses and super-cooled liquids~\cite{Widmer-Cooper_np,rottler_prx,chen_prl,manning_prl,Zylberg_PNAS2017,Flenner_SM2020,Lerner2019JCP, Wang2019SMattenuation,Xu2010EPL}.

Mean field theory \cite{meanfield1,meanfield2} predicts that the density of the low frequency excess modes
$D_\text{exc}(\omega)$ grows as $\omega^{\beta}$ with ${\beta}=2$, while several phenomenological
models~\cite{Buchenau1991,Schober1996,Gurevich_prb2003,Gurarie_prb2003,Kumar2021arxiv} predict ${\beta}=3$ or $4$.
Fluctuating elasticity theory~\cite{Schrimacher_prl2007}
predicts that $D_\text{exc}(\omega)$ scales as $\omega^{d+1}$, which
is the same scaling that the same theory predicts for the sound attenuation coefficient.
An analysis based on a fold stability predicts $D_\text{exc}(\omega)$ scales as $\omega^3$
in glasses approaching marginal stability~\cite{Xu2017prl}.
Other recent theories predict the density of the excess modes scales as
$\omega^4$~\cite{Ikeda2019pre,Stanifer2018pre,Ji2019pre,Bouchbinder2020}.

Simulations are a useful tool to examine low-frequency vibrational modes since characteristics of each mode can
be investigated, but studying finite systems presents
some difficulties. One is that the plane-wave-like modes occur around discrete frequencies, which can be
approximated using Debye theory, but the density of states  does not
exactly follow the Debye prediction. One has to be careful in the calculation of the density of states due to this
discrete nature of the plane-wave-like modes \cite{Bouchbinder2018}.
Simulating two-dimensional (2D) glasses adds another potential difficulty since Mermin-Wagner
\cite{Flenner2015,Vivek2017,Illing2017,Flenner2018}
fluctuations lead to pronounced finite size effects in some static and dynamic properties of two-dimensional
solids.

With increasing system size the glass behaves increasingly as a continuous elastic solid,
and it is expected that there are plane-wave-like modes similar to those of Debye theory.
This observation has lead researchers to distinguish between
plane-wave-like modes and additional modes.
One simple way to do this is to use the participation ratio, which is a measure of how many particles significantly
participate in the mode \cite{Mazzacurati1996,ikeda_pnas}.
A more sophisticated approach is to introduce an order parameter that quantifies the similarity between
a low-frequency mode in an amorphous solid and a plane wave~\cite{ikeda_pnas}.
Although these
two methods are naturally suited for large systems, in principle they can be used for systems of any size.
An alternative approach to distinguish modes of different nature is to study small systems in which
the first plane-wave-like mode (which is found at a frequency close to the one predicted by Debye theory)
is pushed to higher frequencies \cite{lerner_prl2016}. The low-frequency modes found
in these small systems are postulated to be the modes in excess of the Debye prediction.

Mizuno, Shiba, and Ikeda~\cite{ikeda_pnas} used the participation ratio and an order parameter
to separate modes into extended and excess modes in
large, over one-million particle, two- (2D) and three-dimensional (3D) systems. In both dimensions they found that
the density of the modes with large participation ratio obeyed Debye scaling.
In 3D they found that the density of the excess modes, which they determined are quasi-localized, scales as
$D_{loc}(\omega) \sim \omega^4$.

The scaling of the density of excess modes
Mizuno \textit{et al.} found in 3D agrees with the scaling observed previously by studying small systems
\cite{lerner_prl2016}.
Subsequent work by Wang \textit{et al.}~\cite{Wang2019NC}
confirmed that the picture observed by Mizuno \textit{et al.} in 3D is also found in glasses of various
stabilities, up to the stability of laboratory glasses.
Numerical simulations  have demonstrated overwhelmingly the universality of $D(\omega)\sim \omega^4$ scaling
in 3D model glass formers, irrespective of  glass preparation protocols, or  interaction
potentials~\cite{spinglassPRL,lerner_prl2016,ikeda_pnas,ikeda_pre2018,Shimada2018pre,lerner2018jcp,Wang2019NC,
lerner2020pre,Angelani2018,Lerner2020pnas,Bonfanti2020prl,Lopez2020prl,Das2021prl,Shimada2020pre,Ji2020pre,LernerPRE2017}.
In their studies of large 2D systems,
Mizuno \textit{et al.} found very few low frequency modes with small participation ratio or with small
values of plane-wave order parameter.
However, the work of Kapteijns, Bouchbinder, and Lerner~\cite{Lerner2Dprl,Krishnan2D},
who studied small systems, found modes below the first plane-wave-like mode with
density scaling as $\omega^4$.
Notably, unlike in higher dimensions,  Kapteijns \textit{et al.}\ found that
the pre-factor for the $\omega^4$ scaling grew with system size as $[\log{N}]^{5/2}$ in 2D.
For the much larger system studied by Mizuno \textit{et al.}, it might be
expected that there would be a discernible
increase in the density of states over the Debye spectrum, but the logarithmic increase
with system size would make the increase modest.

In this work, we present results for the density of excess modes, $D_\text{exc}(\omega)$, in 2D model glass formers
with different interaction potentials and stability. We used two ways to calculate $D_\text{exc}(\omega)$.
The first method is to subtract off the infinite size system Debye prediction. Except for very
low frequencies, this should allow one to examine how the density of modes in excess of the Debye prediction
changes with frequency, but the discrete nature of the spectrum at low frequencies makes it hard to determine
the low-frequency growth of $D_\text{exc}(\omega)$. We used this concept
and found that $D_\text{exc}(\omega) \sim \omega^2$ in 2D, which differs from previous
observations.  Importantly, $D_\text{exc}(\omega)$ is correlated with the low
frequency scaling of sound attenuation, which resembles the correlation we found between
the density of the excess modes and the sound attenuation in 3D \cite{Wang2019SMattenuation}.
To make a more direct connection with previous results we also studied small systems.  Unlike previous work,
we found a system size and model independent $\omega^3$ scaling of modes far below the first mode predicted
by Debye theory. However, these low-frequency modes are very
rare even in poorly annealed systems and are absent in our very stable systems.

We performed extensive simulation studies of four 2D model glass formers with spherically symmetric interactions:
(I) a system of polydisperse particles interacting via an inverse power law potential $\propto {r}^{-n}$ ($r$ is
the distance of two particles) with $n=12$ (IPL-12)~\cite{berthier_2020nc}; (II) a bidisperse system with
inverse power law potential with $n=10$ (IPL-10)~\cite{lerner2018jcp}; (III)  a bidisperse system with Lennard-Jones
potential (LJ)~\cite{BruningLJ}; (IV)  a bidisperse  system with a harmonic potential (HARM)~\cite{OHern2003}.
Details regarding the four models can be found in the Supplemental Material~\cite{SM}.

We created zero-temperature ($T=0$) glasses by quenching instantaneously  equilibrated liquid configurations at
parent temperatures $T_p$ to $T=0$ using the fast inertial relaxation engine~\cite{fire}.
Equilibrated liquids at very high parent temperatures were obtained by performing molecular dynamic
simulations using LAMMPS~\cite{lammps}. As discussed elsewhere, glasses obtained using this method are
not very stable. To generate stable glasses for the IPL-12 system, we  employed  the swap Monte Carlo
method~\cite{Grigera2001,berthier_prx2017,berthier_prl2016} to prepare equilibrated supercooled liquids
at low $T_p$, down to around $37\%T_g$, where $T_g\approx 0.082$ is
the estimated experimental glass temperature~\cite{berthier_2020nc}.

The normal modes of $T=0$ glasses were obtained by diagonalizing the Hessian matrix using ARPACK~\cite{arpack}
and Intel Math Kernel Library~\cite{mkl}. The density of states is given by
$D(\omega)=\frac{1}{2N-2} \sum_{l=1}^{2N-2} \delta(\omega-\omega_l)$ with $\omega_{l}$ the frequency of mode $l$
and $N$ the number of particles.
In glasses, there are no pure plane-wave modes and the frequencies of the plane-wave-like modes are generally
clustered around the Debye predictions~\cite{Lernerarxiv2021,Bouchbinder2018}. Since Debye theory predicts discrete modes in finite systems,
one has to be careful when calculating the density of states since that calculation requires a division by the bin
size $\delta \omega$. If $\delta \omega$ is not chosen correctly, the density is inaccurate. The calculation of the
cumulative density of states $I(\omega)=\int_{0}^{\omega}D(\omega')d\omega'$ does not suffer from
this issue since it amounts to counting the number of states up to $\omega$ and dividing by the total number
of states. For this reason, we focus on  the cumulative density of states $I(\omega)$.


To obtain the excess modes Mizuno and coworkers~\cite{ikeda_pnas}
defined a threshold of the participation ratio $P_c = 0.01$
to divide plane-wave-like
modes and quasi-localized modes.
They concluded that there are few to no low-frequency quasi-localized modes in poorly annealed 2D
glasses~\cite{ikeda_pnas}. Additionally, the Debye theory accurately predicted the low-frequency
density of states, but there was still a boson peak at higher frequencies.
We attempted to use the participation ratio to separate the modes, but
we found that the scaling behavior of the excess modes in 2D stable IPL-12 model glasses depends strongly on
the choice of $P_c$. This strong dependence makes it impossible to determine the scaling of $I(\omega)$ using
the participation ratio.
Therefore, we utilized a different procedure by subtracting from the cumulative density of states
the Debye prediction~\cite{Schrimacher_prl2007}

\begin{equation}
 I_\text{exc}(\omega)=I(\omega)-I_{D}(\omega),
\label{eq1}
\end{equation}
where $I(\omega)$  is the cumulative density of states of all modes and $I_D(\omega)$ is the Debye
prediction~\cite{kittel},
$I_D(\omega)= A_D\omega^{d}/d$ with  Debye level $A_D$ determined independently from mechanical
moduli~\cite{kittel}.
We note  that this procedure does not take into account that the mode frequencies are discrete for finite systems.

\begin{figure}
\includegraphics[width=0.48\textwidth]{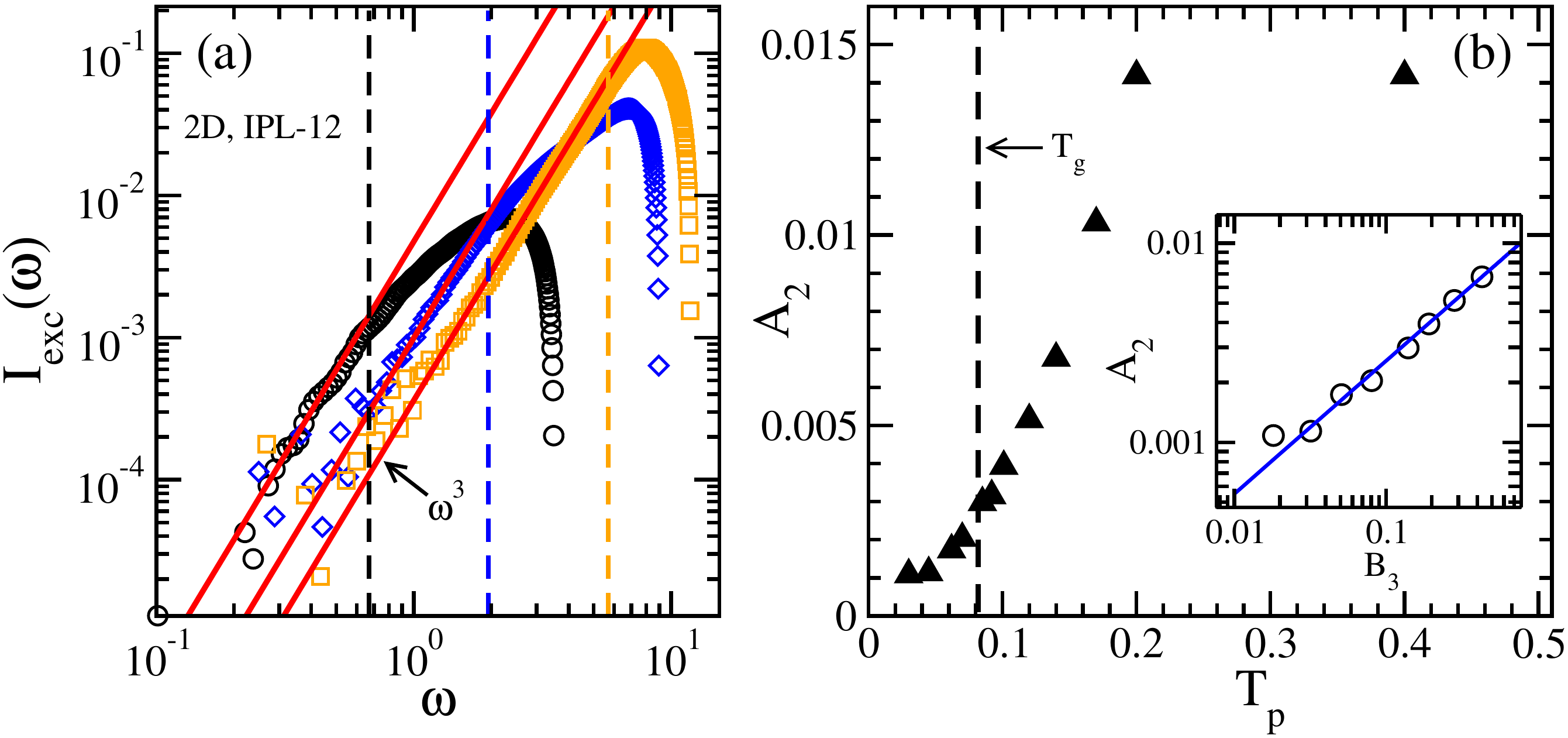}

\caption{\label{fig1} (color online) (a) Cumulative density of states of excess modes
$I_\text{exc}(\omega)=I(\omega)-I_D(\omega)$
at  $T_p=0.400$ (circles), $0.085$ (diamonds), and $0.030$ (squares)  in $N=20000$ system in the 2D IPL-12 model.
$I(\omega)$ is the total cumulative density of states while $I_D(\omega)=1/2A_D\omega^2$ with $A_D$ Debye level.
The red lines are fits to $I_\text{exc}(\omega)=1/3A_2\omega^3$ while  vertical lines indicate boson peak frequencies.
(b) $T_p$ dependence of $A_2$, with the estimated glass transition temperature $T_g$  indicated for reference.
Glass stability increases with decreasing $T_p$.  (Inset)  $A_2$ against the pre-factor $B_3$ in
$\Gamma(\omega)=B_3\omega^3$ with $\Gamma$ the transverse sound attenuation coefficient. Details for
the calculation of $\Gamma$ can be found in Ref.\cite{Wang2019SMattenuation}. The blue line indicates a fit to
$A_2 \sim B_3^{\gamma}$ with $\gamma \approx 2/3$.
 }
\end{figure}

Figure~\ref{fig1}(a) shows excess cumulative density of states $I_\text{exc}(\omega)$  for our 2D IPL-12 model
glasses for $N=20000$ at different parent temperatures $T_p$.  The glass stability increases with decreasing
$T_p$~\cite{Wang2019NC,berthier_2020nc,Lerner2020pnas}. We use parent temperatures
ranging from $T_p=0.400$, which is above
the onset temperature of slow dynamics $T_o=0.250$, down to $T_p=0.030$, which is  below the estimated
laboratory glass temperature $T_g=0.082$ \cite{berthier_2020nc}.
For the lowest frequencies where we can clearly estimate a power law, we find that
$I_\text{exc}(\omega)$ scales as $\omega^3$, $I_\text{exc}(\omega)\simeq A_2\omega^3/3$,
which suggests that $D_\text{exc}(\omega)\simeq A_2\omega^2$.
We find that this scaling continues up to around the Ioffe-Regal limit, which is around the boson peak frequency~\cite{tanaka_nm2008},
irrespective  of  the glass's stability, see Fig.~\ref{fig1}(b).
We find that the coefficient quantifying the magnitude of the excess modes density, $A_2$, is stability-dependent.
$A_2$ is nearly constant for larger $T_p$, but decreases by a factor of 13
for our lowest $T_p$. This indicates that there are fewer excess modes for increasingly stable 2D glasses,
which is consistent with observations for 3D glasses~\cite{Wang2019NC,Lerner2020pnas,Ji2020pre}.

Previous work~\cite{Wang2019SMattenuation,Lerner2019JCP,Schrimacher_prl2007} found a connection between
sound attenuation and  density of states of excess low-frequency modes. Inspired by this work we examined
whether $I_\text{exc}(\omega)$  is related to sound attenuation in 2D.
The frequency dependence of the transverse sound attenuation coefficient $\Gamma(\omega)$ in 2D glasses follows
the Rayleigh scattering scaling as $\Gamma(\omega) =B_3\omega^3$
and $B_3$ decreases with increasing glass stability~\cite{Lerner2019JCP}. Here, we study the relation between
$B_3$ and $A_2$, see the inset to Fig.~\ref{fig1}b.
We find $A_2 \sim B_3^{\gamma}$ with $\gamma\approx 2/3$, and thus we
establish that in 2D the excess modes
density is related to sound attenuation, which is consistent with our result for 3D glasses.

\begin{figure}
\includegraphics[width=0.42\textwidth]{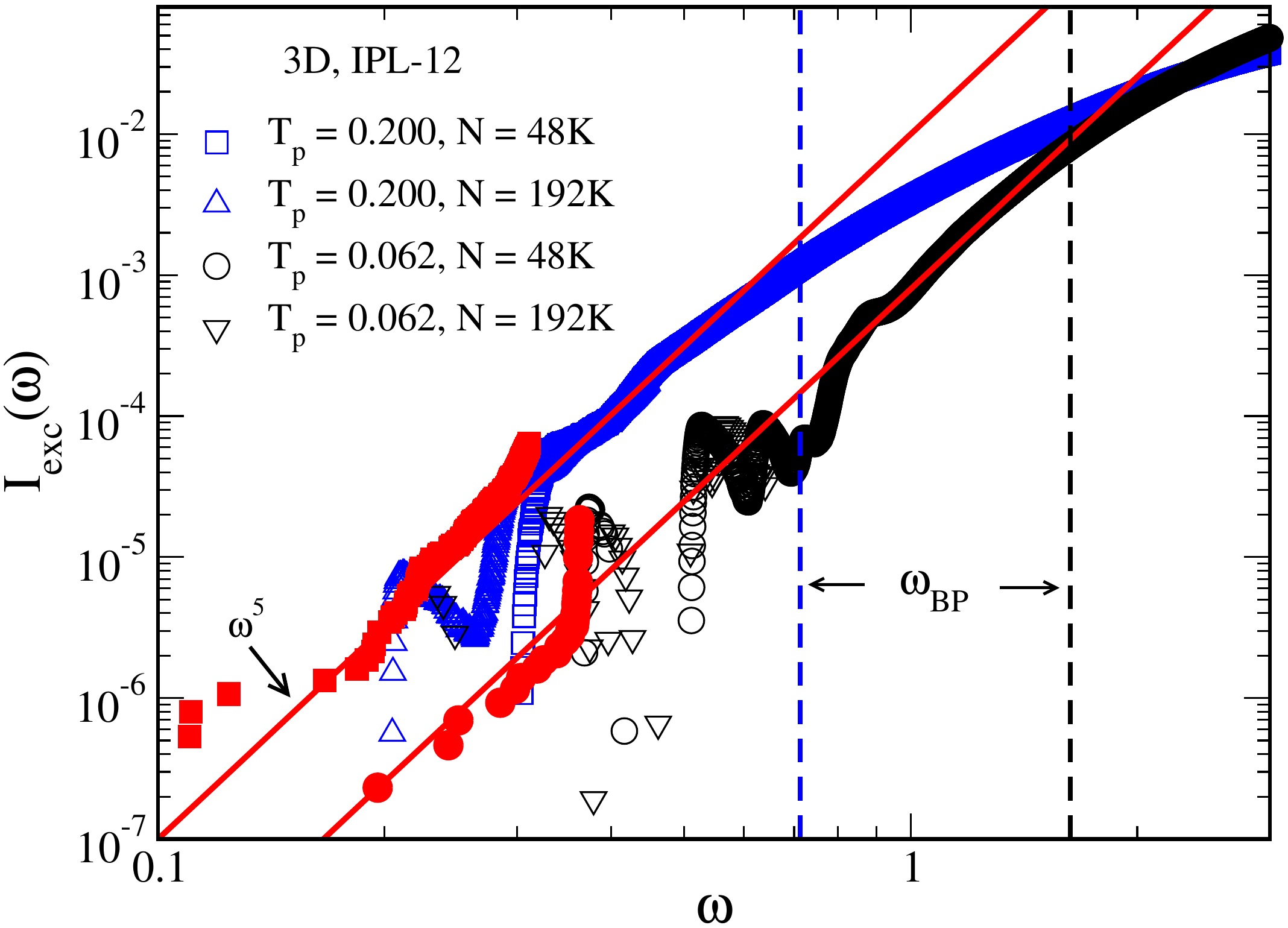}

\caption{\label{fig2} (color online)  Cumulative density of states of excess modes
$I_\text{exc}(\omega)=I(\omega)-I_D(\omega)$ in the 3D IPL-12 model glasses, with
the boson peak frequency indicated.  $I(\omega)$ is the total cumulative density of states and
$I_D(\omega)=A_D\omega^3/3$ with $A_D$ being the Debye level.
Red filled squares and circles represent $I(\omega)$ at frequencies below around the first Debye
frequency at $T_p=0.200$ and $T_p=0.062$,
respectively, in the $N=48000$ system. The red lines correspond to $I_\text{exc}(\omega)\sim \omega^5$. }
\end{figure}

Since the method introduced in this Letter is different from methods used before by us and others, we checked
what results it produces if used for 3D glasses where it has been firmly established that
$I_\text{exc}(\omega) \sim \omega^5$.
In Fig.~\ref{fig2} we show  $I_\text{exc}(\omega)$ in 3D IPL-12 glasses for two stabilities,
a very poorly annealed glass with $T_p=0.200$ and a very stable glass with $T_p=0.062$.
These are the same  glasses examined in Ref.~\cite{Wang2019NC}.  We find that $I_\text{exc}(\omega) \sim \omega^5$
up to a frequency close to the boson peak for both glasses, which indicates the resulting scaling of
$I_\text{exc}(\omega)$ determined using Eq.~\ref{eq1} is consistent with that of $I_\text{exc}(\omega)$
calculated with previously used procedures~\cite{Wang2019NC,ikeda_pnas,ikeda_pre2018}.
Additionally, these results suggest that in 3D
the end of the $\omega^5$ scaling of $I_\text{exc}(\omega)$, $\omega_g$, is around the boson peak frequency.
We note that our procedure cannot be used for frequencies below the lowest frequency plane wave mode predicted by
the Debye theory. More importantly, it will only reveal the proper scaling if there is a near continuum
of modes \cite{Bouchbinder2018}.

Debye theory predicts that the lowest normal mode frequency increases with decreasing $N$. It has been argued
that the scaling of the excess modes could be obtained from the low frequency density of states for small systems
since the frequency of the plane-wave-like modes is pushed to high frequencies. Using this reasoning,
$I_\text{exc}(\omega)$ can be determined by calculating
the total  cumulative   density of states  $I(\omega)$ for frequencies below the first predicted Debye mode frequency.
One may expect $I(\omega)=I_\text{exc}(\omega)$ for low frequencies if the excess modes were independent of
the plane--wave like modes, which we found for 3D glasses with different stabilities, see Fig.~\ref{fig2}.
The low-frequency  tail of $I(\omega)$
is well described by a power law $I_\text{exc}(\omega)\sim\omega^5$ for $T_p=0.200$ and $T_p=0.062$.
However, in 2D glasses much below the first Debye frequency we find $I(\omega)\sim\omega^4$, which
suggests that $D_{exc}(\omega) \sim \omega^3$.
Previous studies reported  $D_\text{exc}(\omega)\sim \omega^4$~\cite{Lerner2Dprl,Krishnan2D},
which would imply that $I_\text{exc}(\omega) \sim \omega^5$. To make sure this observation is model independent,
we calculated $I(\omega)$ for small systems at frequencies much lower than the first Debye frequency in different
models of glass formers.

\begin{figure}
\includegraphics[width=0.45\textwidth]{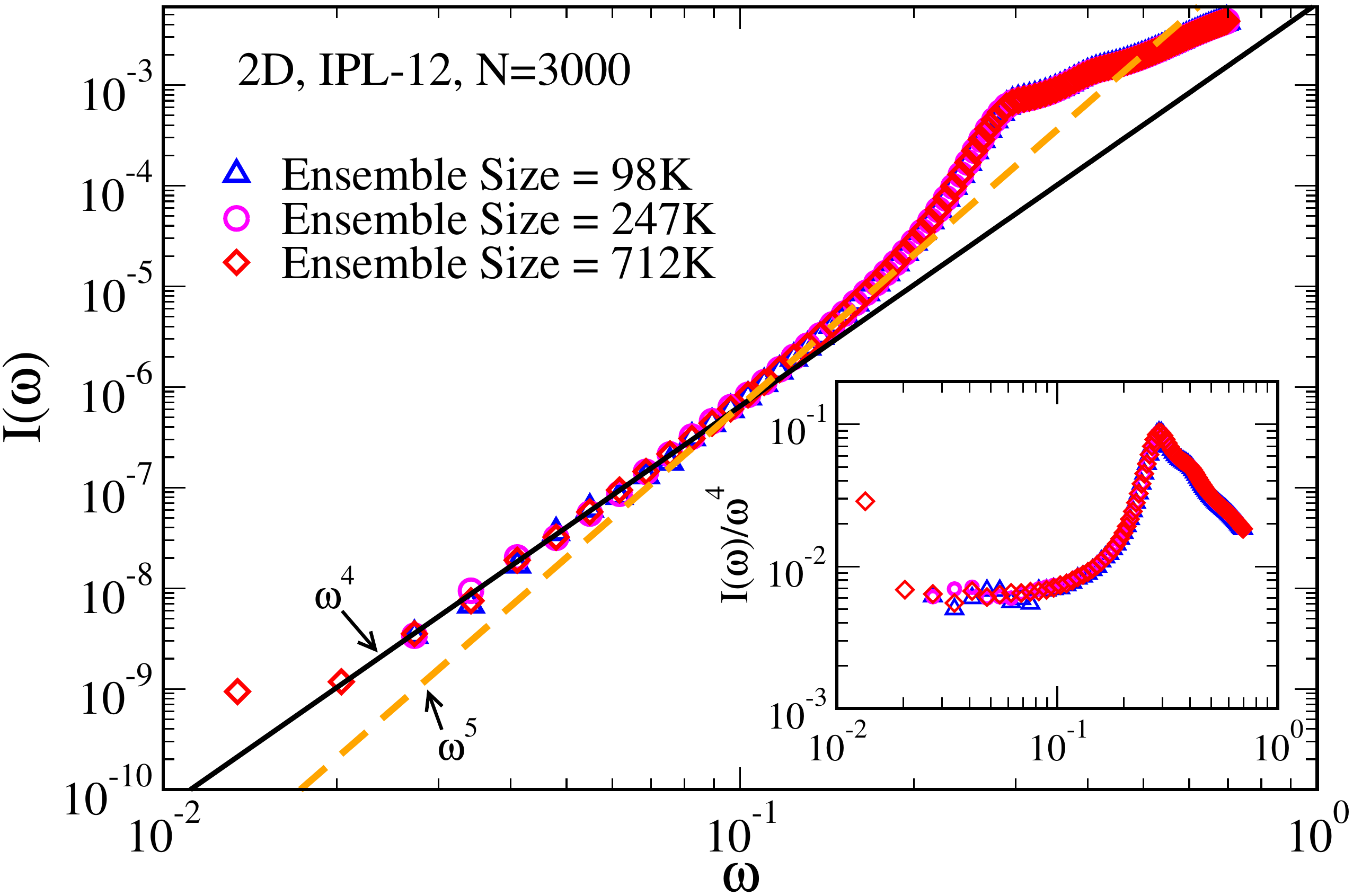}

\caption{\label{fig3} (color online) The total cumulative density of states  $I(\omega)$  for $N=3000$ system
at $T_p = 0.400$
with different ensemble sizes  in the 2D IPL-12 model. The solid  and dashed lines represent power laws of
$\omega^4$ and $\omega^5$ respectively. For reference, the lowest Debye mode frequency is about $0.261$.
(Inset) The same data plotted as $I(\omega)/\omega^4$ vs. $\omega$.  }
\end{figure}

In Fig.~\ref{fig3} we  show the total cumulative density of states $I(\omega)$ for $N=3000$ system in the 2D
IPL-12 model. There is a range of frequencies below the lowest Debye mode frequency ($\approx0.261$) that
is well described by $I(\omega)\sim\omega^4$.
To check  the quartic scaling, we examined  $I(\omega)/\omega^4$, which is shown in the inset to Fig.~\ref{fig3},
and we find that there is a low-frequency plateau.
Previous results suggest that $I(\omega)\sim \omega^5$  at frequencies much below the first Debye mode in 2D
glasses~\cite{Lerner2Dprl,Krishnan2D}, and that $I(\omega)$ should be system size dependent.
However,  we find that the  $\omega^5$ scaling is only valid for an intermediate-frequency regime below
the peak of $I(\omega)$, and appears to be only a transition between
the low frequency scaling and the change of the scaling due to the emergence of plane-wave-like modes.
We also find that $I(\omega)$ is system size independent at frequencies much
lower than the first frequency predicted by Debye theory.

\begin{figure}
\includegraphics[width=0.46\textwidth]{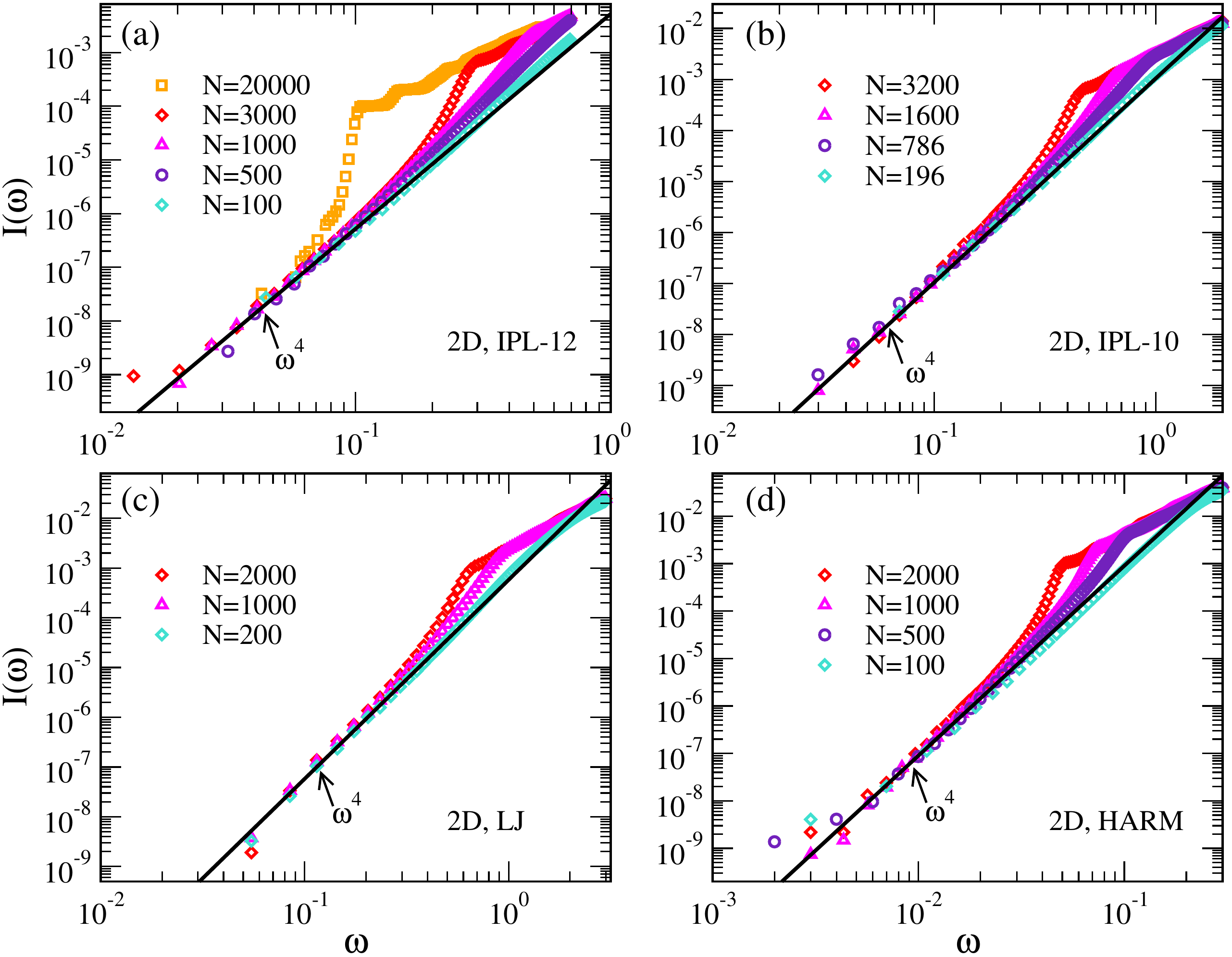}

\caption{\label{fig4} (color online)   The total cumulative density of states $I(\omega)$  for different
system sizes in (a) the 2D IPL-12 model, (b)  2D IPL-10 model,  (c)  2D LJ model and (d)  2D HARM model.
In each panel, the solid line corresponds to the power law of $\omega^4$ and represents the low frequency data
very well for all model glass formers examined. We plotted the same data as $I(\omega)/\omega^4$ against $\omega$ in the Supplemental Material \cite{SM}.}
\end{figure}

There are very few modes that contribute to the low frequency $\omega^4$ scaling of
$I(\omega)$ for our least stable 2D glass.
For example,  on average there is  only one mode  lying in the low-frequency $I(\omega)\sim \omega^4$
regime every one hundred configurations for the $N=3000$ system.
The lower the frequency we want to examine, the larger the ensemble size $N_{En}$ (number of configurations) we need.
However,  we do not observe $N_{En}$  dependence of  the quartic scaling regime in $N=3000$ system when
$N_{En}$ ranges from around 0.1 million to  our maximum 0.71 million examined, see Fig.~\ref{fig3}.
The same conclusion can also be drawn in our study of the $N=1000$ system where the maximum $N_{En}$ is
around 2.2 million \cite{SM}. In addition, we checked that the previously reported $\omega^5$ scaling in some systems
is due to ensemble size not being large enough, which hinders the observation of the $\omega^4$
scaling at much lower frequencies. We also find that the $\omega^5$ scaling regime vanishes for very small systems,
Fig.~\ref{fig4}. Since the number of these low-frequency modes decreases with increasing stability, we could
not examine the stability dependence of these modes. However, we do not exclude the possibility that  thermal relaxation can change the scaling of these low-frequency modes~\cite{LernerPRE2017}.

We find that this low-frequency quartic scaling of $I(\omega)$ in 2D is universal, \textit{i.e.} it does
not depend on the model glass former, see Fig.~\ref{fig4}. In addition, there are
common features shared by each model. First, the pre-factor of the quartic scaling
does not depend on system size. This conclusion is different than the conclusion of Ref.~\cite{Lerner2Dprl}, namely
that $I_\text{exc}(\omega)=A_4\omega^5/5$ and the pre-factor $A_4$ grows as $(\log N)^{5/2}$, \textit{i.e.} there
should be more excess modes in a larger system.
Second, the low-frequency scaling $I(\omega)\sim\omega^4$ works up to  a larger frequency with decreasing system size.
It deserves further study to check whether the  upper frequency for this scaling correlates with any known
characteristic frequency. Figures~\ref{fig3} and~\ref{fig4}, therefore, demonstrate that
$I(\omega)\sim\omega^4$  at frequencies much lower than the first frequency predicted by Debye theory in 2D glasses
and  the pre-factor for this power law shows no system size dependence.

We utilized two methods to examine the excess density of states in 2D glasses.
In large systems,  we find evidence that the excess density of states scale as $\omega^2$.
We find that the pre-factor $A_2$  of this scaling law correlates with the sound attenuation
coefficient. However, in small systems, where the frequencies of plane-wave-like modes are pushed higher,
we find that the modes below the lowest Debye frequency have density of states scaling as $\omega^3$,
with a system size-independent pre-factor. This inconsistent behavior is not found in 3D glasses
using the same analysis.

Our results leave several open questions. First, why is the scaling of excess modes different above and
below the first mode predicted by Debye theory?
One possibility is that our systems are not large enough to accurately determine $I_\text{exc}(\omega)$
by subtracting off the Debye contribution at low frequencies.
However, we do find an overlapping
frequency range where we find that the excess density of states scales as $\omega^2$ when we subtract
off the Debye contribution and where the excess density of states scales as $\omega^3$ for very small systems,
see Fig. 3 of the Supplemental Material \cite{SM}.
Thus, it seems that the presence of plane waves influences the
scaling of the excess density of states in 2D glasses.

There may also be a gap in the excess density of states
and the $\omega^2$ and the $\omega^3$ scaling does not extend to $\omega = 0$,
which would be consistent with the conclusions of Ref.~\cite{ikeda_pnas}.
It is very difficult to numerically test these possibilities
with the current computer power since the $\omega^2$ and $\omega^3$ scaling represents
very few modes at small $\omega$.
Future related theoretical work may shed some light on this issue.

Third, why is the excess density of states different from that of 3D glasses? Fluctuation elasticity theories
predict that $D_\text{exc}(\omega)$ depends on spatial dimension as $ \omega^{d+1}$~\cite{Schrimacher_prl2007}.
Thus, the predicted  $D_\text{exc}(\omega)\sim \omega^4$ in 3D glasses is consistent with 3D numerical observations.
The predicted scaling  of $D_\text{exc}(\omega)\sim \omega^3$ of 2D glasses is consistent with what we find in small
systems. However, it has been shown that $D_{\text{exc}}(\omega) \sim \omega^4$  in 2D glasses \cite{Lerner2Dprl} in contrast to
the fluctuating elasticity theory prediction. We notice that the  dynamics of 2D and 3D glass-forming liquids was
reported to be fundamentally
different~\cite{Flenner2015}. It may be also interesting to probe whether there is any connection between
the difference in dynamics between 2D and 3D glass-forming liquids and the difference in the density of states of
excess modes between 2D and 3D glasses.

Finally, it is possible that the upper frequency cutoff of the low frequency scaling, $\omega_g$, is below the frequency
range where we found $\omega^2$ scaling of $D_\text{exc}(\omega)$ in  large 2D glasses.
If this were the case, it would differ from our finding in 3D that $\omega_g$ is around the boson peak frequency.
We leave it for future work to examine what determines $\omega_g$ and if it depends on dimension.

We wish to thank Andrea Ninarello for generously providing  equilibrated configurations at very low parent
temperatures and E. Lerner for comments on the manuscript.
L. W. acknowledges the support from  National Natural Science Foundation of China (No. 12004001),
Anhui Province (No. S020218016), Hefei City (No. Z020132009), and  Anhui University (Start-up
fund). E.F. and G.S.  acknowledge the support from  NSF Grant CHE-1800282.

\begin{widetext}
\newpage
\

\includepdf[pages=-]{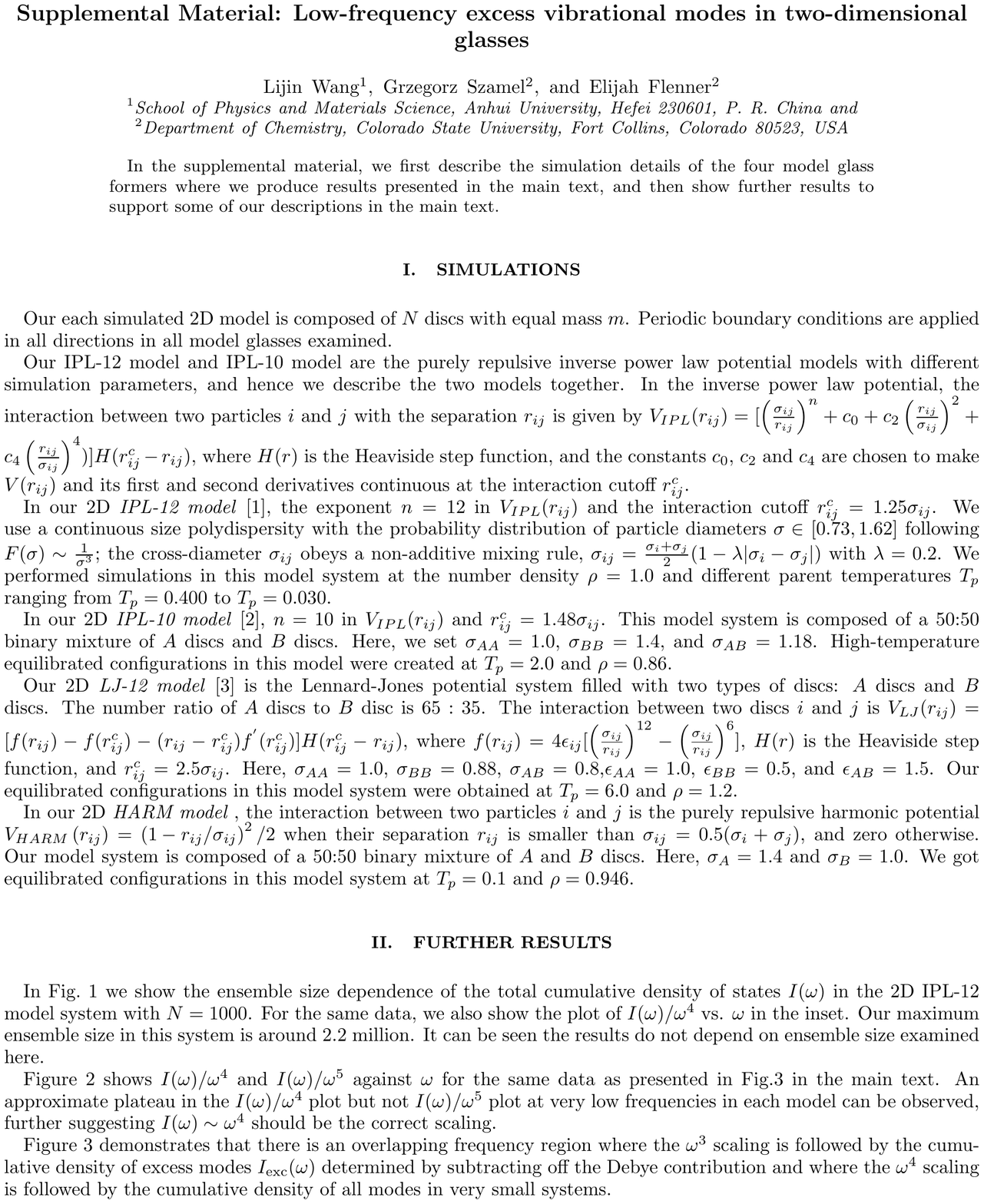}
\end{widetext}

\end{document}